*Lunar-like silicate material forms the Earth quasi-satellite (469219) 2016 HO$_3$ Kamo`oalewa*

Benjamin N.L. Sharkey[1,*], Vishnu Reddy[1], Renu Malhotra[1], Audrey Thirouin[2], Olga Kuhn[3], Albert Conrad[3], Barry Rothberg[3,4], Juan A. Sanchez[5], David Thompson[3], Christian Veillet[3]

*sharkey@lpl.arizona.edu

[1]Lunar and Planetary Laboratory
University of Arizona
1629 E University Blvd
Tucson, AZ 85721, USA

[2]Lowell Observatory
1400 W Mars Hill Road
Flagstaff, AZ 86001, USA

[3]Large Binocular Telescope Observatory
University of Arizona
933 N. Cherry Ave, Room 552
Tucson, AZ 85721, USA

[4]George Mason University
Department of Physics and Astronomy
MS3F3
4400 University Drive,
Fairfax, VA 22030, USA

[5]Planetary Science Institute
1700 E Fort Lowell Suite 106
Tucson, AZ 85719, USA

**Abstract**

Little is known about Earth quasi-satellites, a class of near-Earth small solar system bodies that orbit the sun but remain close to the Earth, because they are faint and difficult to observe. Here we use the Large Binocular Telescope (LBT) and the Lowell Discovery Telescope (LDT) to conduct a comprehensive physical characterization of quasi-satellite (469219) Kamo`oalewa and assess its affinity with other groups of near-Earth objects. We find that (469219) Kamo`oalewa rotates with a period of 28.3 (+1.8/-1.3) minutes and displays a reddened reflectance spectrum from 0.4-2.2 microns. This spectrum is indicative of a silicate-based composition, but with reddening beyond what is typically seen amongst asteroids in the inner solar system. We compare the spectrum to those of several material analogs and conclude that the best match is with lunar-like silicates. This interpretation implies extensive space weathering and raises the prospect that Kamo`oalewa could comprise lunar material.

**Introduction**

Near-Earth object (NEO) (469219) Kamo`oalewa (provisional designation 2016 $HO_3$) is the most stable of the five known quasi-satellites of the Earth, with a dynamical lifetime of a few hundred years [1]. As a quasi-satellite, the orbit of Kamo`oalewa is very Earth-like, with semi-major axis within 0.001 au of Earth's, a low eccentricity of just ~0.1, and a modest inclination of about 8 degrees to the ecliptic [2] and it is a frequently proposed target for spacecraft study [3-5]. As it orbits the Sun with a ~1 year orbital period, it takes a quasi-satellite path relative to Earth, that is, it makes retrograde loops around Earth with a ~1 year period but well beyond Earth's Hill sphere [6]. Study of this class of objects began with the initial discovery of (164207) 2004 $GU_9$ [7-9]. Physical characterization of the quasi-satellite population has been lacking due to challenging observing geometry and short residence time in near-Earth space. Uniquely, Kamo`oalewa is favorably placed for observations once a year around April when it becomes bright enough (with visual magnitude V<23.0mag) to be characterized by large telescopes on Earth. These regular observing windows allow for continued study, unlike temporarily captured minimoons such as 2020 $CD_3$ that require dedicated observing campaigns during a single apparition shortly after their discovery [10-11].

**Results and Discussion**

We used the LBT to obtain broadband color photometry and visible spectra on UT 14 April 2017. The Multi-Object Double Spectrograph (MODS) instrument was used in imaging and long-slit spectroscopic modes [12] to carry out our observations. These observations provided an initial assessment of the quasi-satellite's taxonomic class as being consistent with S- and L-type silicate asteroids [13] and rotation period, determined to be < 30 minutes upon the object's discovery [14].

Follow-up broadband photometry was collected on UT 18 April 2017 at LDT in the VR filter using the Large Monolithic Imager [15]. Exposure times varied from 200s to 300s. The goal of these observations was to collect high signal-to-noise ratio data, which combined with the LBT observations provided a longer baseline in time to derive the object's rotational lightcurve and so improve the period determination.

Using established data calibration and reduction routines [16], we find Kamo`oalewa's rotational period to be 28.3 (+1.8/-1.3) minutes (99th percentile error limits), consistent with the findings upon its discovery [14]. The peak-to-peak lightcurve amplitude is 1.07±0.05mag based on a Fourier fit, corresponding to an axis ratio of *a/b*≥1.53 [17]. The lightcurve is plotted in Figure 1, which also provides a comparison of Kamo`oalewa's spin period with that of known NEOs [18], showing that its rotation state is typical of similarly sized NEOs. The population of NEOs with H > 21.0 mag rotate fast enough to imply internal cohesion [19], illustrated by curves indicating the maximum spin speed of an object at a given size given a range of assumed object bulk densities ($\rho_b$) and tensile strength coefficients ($\kappa$). Previous study of plausible shape and rotation models for Kamo`oalewa found that Kamo`oalewa's surface may retain grains smaller than ~1 mm-1 cm in size [20].

Our observed reflectance spectrum at visible wavelengths (0.4-0.95 μm) is consistent with many types of silicate materials which are common among near-Earth asteroids, particularly S- and L-taxonomic classes, but is not diagnostic of any single compositional category (Figure 2, note that all spectra plot very closely at wavelengths below 0.8 μm). Follow-up spectroscopic observations in the near-infrared (1.0-1.3 μm) obtained in 2019 confirmed the presence of a 1.0-micron absorption band, due to the minerals olivine and/or pyroxene, but superimposed on a steep red continuum slope. This slope was confirmed by measuring broadband zJH photometric colors on UT 7 March 2021 with the LBT, and JK colors on UT 11 April 2021.

We used curve matching to constrain the surface composition and identify possible analog materials for Kamo`oalewa. The spectral slope of Kamo`oalewa is redder (higher reflectance at increasing wavelength) than typical S-type asteroid spectra as defined by the Bus-Demeo taxonomy. As an example, Figure 2 shows the spectrum of Kamo`oalewa with that of Sw-type asteroid (63) Ausonia [21-22]. Sw-types exhibit the same characteristics as S-types but have redder slopes in the infrared [23]. The spectral slope of Kamo`oalewa is even redder than the slope of these objects. Reddening of spectral slope can be due to a range of factors including phase angle (Sun-Target-Observer angle), particle size of the regolith (increasing grain size causes bluer spectral slope), metal content (increasing metal causes redder spectral slope), and lunar-style space weathering (increasing space weathering causes redder spectral slope and weaker absorption bands).

Laboratory measurements of LL ordinary chondrite meteorites, which are derived from S-type asteroids, have shown maximal spectral slope changes by up to 30%/1 μm due to phase angle variations of ~120 degrees [24]. The NIR spectrum of Kamo`oalewa was observed at a phase angle of 43.2° and its slope (measured by the online Bus-DeMeo taxonomy tool [25]) of 89%/1μm is too steep to be explained by just phase angle effects. The grain size of the regolith also influences spectral slope and band depth. Typically for silicates, the band depth and spectral slope decrease (becoming blue-sloped) as grain size approaches millimeters to centimeters. In the case of Kamo`oalewa, the band depth is weak, but the spectral slope is red—not blue as would be expected from grain size effects alone. Hence, grain size effects are insufficient to reconcile the observed spectral properties of Kamo`oalewa with a typical S-type composition with a more neutral reflectance spectrum.

The choice of characterizing the spectral slope of Kamo`oalewa using the standard Bus-DeMeo taxonomic tool [23] was made to enable general comparisons with the NEA population. This method involves fitting a straight line to a well-sampled spectrum from 0.45-2.45 µm. Since our spectrum of Kamo`oalewa is not uniformly sampled in wavelength, the calculation of the slope was performed by first linearly interpolating the spectrum. To assess the uncertainty of the slope measurement, which is dominated by the errors in the infrared reflectance, the slope measurement was repeated after varying the data at wavelengths greater than 1.0 µm uniformly by either $\pm 1\sigma$ (prior to interpolation). While this is an overestimate of the error (as it assumes the photometric errors are correlated), it provides a transparent look at the variability inherent to our unevenly sampled spectrum. This method returns a range of slope values from 76-101%/1µm. Figure 3 shows the range of slopes plotted with those of near-Earth asteroids compiled by the MITHNEOS survey [26]. The one object with an overall spectral slope which overlaps the range for Kamo`oalewa is (4142) Dersu-Uzala, a Mars-crossing asteroid that was originally classified as an A-type [27] and whose 0.4-1.0 micron spectral slope and 1.0 micron band differs from Kamo`oalewa.

Metal-silicate mixtures have been shown to have spectra similar to Kamo`oalewa with steep spectral slope and a weaker olivine or pyroxene band. Such materials would be analogous to pallasite, mesosiderite, or CB chondrite meteorites. Spectral reddening has been shown to occur in mixtures where the metal content is >50% [28]. We compared the spectral slope and band depth of Kamo`oalewa with metal+olivine and metal+pyroxene mixtures. To illustrate this hypothesis, reflectance measurements were collected of the iron meteorite Gibeon and mesosiderite Vaca Muerta, with a linear combination of these two endmembers plotted in Figure 2. This approach uses only one source of meteoritic metal for comparison, but we note that the spectrum of the Gibeon sample we use is iron-rich, and therefore among the reddest possible analogs amongst meteoritic metal sources. When mixed with spectrally neutral silicates to fit the 1.0-micron band, the spectral slope becomes even less red. Therefore, we conclude generally that mixtures of meteoritic metal with unweathered silicates do not provide a suitable match to Kamo`oalewa's spectrum.

Lunar style space weathering has been demonstrated to reduce the albedo and band depth and increase the spectral slope of a reflectance spectrum [29-30]. We compared the spectrum of Kamo`oalewa to Apollo lunar samples from the RELAB database [31-34]. A total of 19 available samples were used to identify the best spectral match to Kamo`oalewa. Based on our search, the 20-45 µm grains from Apollo 14 Lunar sample #14163 provides a close match amongst surveyed materials to our reflectance measurements of Kamo`oalewa (Fig. 2). However, a variety of slopes is present among the lunar soil sample database, including higher and lower slopes. This shows the inherent uncertainty in separating specific compositional hypotheses based on our low-resolution data. We note that spectral slope matches to fine-grained lunar materials does not necessarily imply a fine-grained regolith on the surface of Kamo`oalewa, as recent in-situ observations by the Chang'e-4 rover of a rock >20cm in size found it to highly reddened [35]. The spectrum of this rock traces closely with Apollo sample 14163 at wavelengths < 1.75 microns, but rolls over to neutral reflectance from 1.75-2.5 microns.

Regardless of its origin, we show via Figures 2 and 3 that the overall spectrum of Kamo`oalewa is inconsistent with typical near-Earth asteroid taxonomies and requires additional explanation

(e.g. high metal content and/or extreme Lunar style space weathering) for its reflectance properties. We note that S-type asteroids often have higher visible albedos than the lunar samples we compared to. For example, the 20-45 µm grains of Apollo 14 sample #14163 has a visible albedo of ~0.1 (although this is dependent on grain size), while S-type asteroid Ausonia has an albedo of ~0.16 [36]. Adopting a range of albedos from 0.10-0.16 provides an effective diameter for Kamo`oalewa (H=24.3mag) in the range from D=58-46 m. Direct measurements of Kamo`oalewa's albedo (or size) coupled with knowledge of the grain size distribution of its surface would therefore play a useful role in discriminating between compositional hypotheses.

Planetary perturbations make Kamo`oalewa's annual retrograde loops around the Earth slightly variable in amplitude over a longer period of about 40 years. Numerical propagation of its orbit over decadal timescales in the past and in the future finds that these nearly-periodic variations remain small (its annual epicyclic path relative to Earth's orbit remains at a geocentric distance of 10-30 $r_{H\oplus}$, where $r_{H\oplus} \approx 1.5 \times 10^6$ km is the radius of Earth's Hill sphere), but larger changes occur over timescales of centuries, as seen in Figure 4. The longer-term orbit propagation finds that the quasi-satellite motion began approximately 100 years ago and will last for about 300 years in the future. Prior to a century ago (and beyond ~300 years in the future), Kamo`oalewa's annual epicyclic loops slowly drift away from Earth, tracking a horseshoe-shaped path in a frame co-rotating with Earth around the Sun; in this state, its semi-major axis alternately falls below and above Earth's by about 0.005 au, and it approaches Earth closely (within ~10 $r_{H\oplus}$) only every ~130 years. In Fig. 4, the quasi-satellite state is indicated by the green track, which shows the small amplitude librations of the semi-major axis about 1 au (Fig. 4a) and of the mean longitude about Earth's mean longitude (Fig. 4b) and the small geocentric distance (Fig. 4c); the horseshoe state is indicated by the black track which shows that the semi-major axis alternately falls below and above Earth's by about 0.005 au (Fig. 4a) and the mean longitude relative to Earth's has large amplitude librations about 180 degrees (Fig. 4b) and the geocentric distance varies up to about 2 au (Fig. 4c). The retrograde quasi-satellite orbits, the horseshoe orbits, and transitions between these two types are known dynamical behaviors near the 1:1 mean motion resonance in the three-body problem [37]. The current errors in the orbital parameters grow beyond the ± 500 year timeframe, precluding confidence in its longer term orbital path. However, numerical sampling of its possible longer term orbital path has been carried out with models of various degrees of fidelity [1,38-41]. These studies commonly indicate that Kamo`oalewa remains in an Earth-like orbit, exhibiting intermittent transitions between horseshoe and quasi-satellite motion, on ~$10^5$-$10^6$-year timescales, but this state is unlikely to persist over timescales comparable to the age of the Earth and of the solar system.

## Conclusions

The natural question that arises is: what is Kamo`oalewa's origin? The answers are speculative. One possibility is that it was captured in its Earth-like orbit from the general population of NEOs. Its low eccentricity and inclination are, however, rather atypical of such captured co-orbital states found in numerical simulations [42]. Another possibility is that Kamo`oalewa originates from an as-yet undiscovered quasi-stable population of Earth's Trojan asteroids orbiting near Earth's L4 and L5 Lagrange points [43-44]. This hypothesis can be tested in future deeper and wider observational surveys of the Earth-Sun Trojan regions, supplemented with theoretical investigation of dynamical pathways between Earth Trojans and quasi-satellites. A third possibility is that Kamo`oalewa originates in the Earth-Moon system, perhaps as impact ejecta from the lunar surface [45] or as a fragment of a parent NEO's tidal or rotational break up during a close encounter with Earth-Moon [46]. Three NEOs, 2020 $PN_1$, 2020 $PP_1$ and 2020 $KZ_2$, have been identified as having orbital parameters cluster near Kamo`oalewa's closely enough that they may be break-up companions [47]. The reflectance spectrum of Kamo`oalewa (as reported in the present work) lends support to the lunar ejecta hypothesis. An origin near or within the Earth-Moon system is further supported by the low value of the relative velocity, v ≈ 2-5 km/s, of Kamo`oalewa during its close approaches to Earth-Moon, whereas NEOs have larger relative velocities at close approaches, with an average v ≈ 20 km/s [48-49].

## **Methods**

Data collected using the twin MODS instrument (one per mirror) on LBT used an exposure time of 120s for all images, which were taken using Sloan i' and z' filters, and exposure times of 900s for the spectral frames. The G2V star HIP 52192 was used as the solar analog to convert raw spectral fluxes to reflectance.

The data calibration and reduction routines described in [16] were applied to the visible photometric data set to determine the object's rotational period. After the light-time correction, we performed a search for periodicity using the Lomb periodogram technique [50]. We find one main peak with a confidence level >99.9% located at ~102.4 cycles/day (i.e, 0.234h) and corresponding to a single-peaked lightcurve (Figure 5). However, the measured lightcurve presents a strong asymmetry with one minimum deeper than the other one by about 0.65 mag. Therefore, we conclude that the double-peaked lightcurve corresponds to the object's full rotation.

Near-infrared spectra were collected as a series of 300s exposures nodded in an AB pattern, totalling 110 minutes of exposure time. Data were processed, including sky subtraction and wavelength fitting, via the Flame pipeline [51] using SAO 120107 as a solar analog. Extraction was performed using an optimized extraction technique [52] implemented in python [53]. As a validation of our techniques, we observed the main-belt asteroid (26) Proserpina in August 2019 using the NASA Infrared Telescope Facility and the twin LUCI (LBT Utility Camera in the Infrared) spectrographs on LBT. We found close agreement between both measurements using the same processing on the LBT data as implemented for Kamo`oalewa (Figure 6).

Color measurements were obtained by measuring target fluxes through standard aperture photometry, relative to the G2V solar analog star GSPC P330-E [54]. On UT 7 March 2021, both the target and the solar analog were observed at airmass <1.01. On UT 11 April 2021, the target was observed over an airmass range of 1.07-1.16, and the solar analog was observed at airmass <1.01. To ensure that the time variability of the target's brightness did not affect color measurements, images were obtained in two filters (zH, JH, and JK) simultaneously to derive the reflectance ratios between each filter combination. This analysis made use of the Photutils package in Astropy [55-57].

**Acknowledgments**

B.N.L. Sharkey and V. Reddy's work presented in this paper is supported by NASA Near-Earth Object Observations Program Grant NNX17AJ19G (PI: Reddy). This paper uses data taken with the MODS spectrographs built with funding from NSF grant AST-9987045 and the NSF Telescope System Instrumentation Program (TSIP), with additional funds from the Ohio Board of Regents and the Ohio State University Office of Research. The LBT is an international collaboration among institutions in the United States, Italy and Germany. LBT Corporation partners are: The University of Arizona on behalf of the Arizona Board of Regents; Istituto Nazionale di Astrofisica, Italy; LBT Beteiligungsgesellschaft, Germany, representing the Max-Planck Society, The Leibniz Institute for Astrophysics Potsdam, and Heidelberg University; The Ohio State University, representing OSU, University of Notre Dame, University of Minnesota and University of Virginia.

Our results also made use of Lowell Observatory's Discovery Telescope (LDT). Lowell operates the LDT in partnership with Boston University, Northern Arizona University, the University of Maryland, Yale University and the University of Toledo. The Large Monolithic Imager was built by Lowell Observatory using funds from the National Science Foundation (AST-1005313). We acknowledge the LDT operator, Jason Sanborn. Audrey Thirouin acknowledges Lowell Observatory funding.

Taxonomic type results presented in this work were determined, in whole or in part, using a Bus-DeMeo Taxonomy Classification Web tool by Stephen M. Slivan, developed at MIT with the support of National Science Foundation Grant 0506716 and NASA Grant NAG5-12355.

Renu Malhotra acknowledges funding from NASA (grants 80NSSC18K0397 and 80NSSC19K0785) and the Marshall Foundation of Tucson, AZ.

**Data availability**

The processed data products reported in each figure, as well as the raw telescope data and calibration files necessary to support independent processing, are available as a single archive at the following DOI: 10.5281/zenodo.5542337

Data from Binzel et al. (2019) can be accessed via:

https://data.mendeley.com/datasets/j96dmmsxrg/1

**Code availability**

Processing scripts are freely available to download via the references provided in the methods section. Python scripts written to process the near-infrared data are available upon request to the corresponding author (https://github.com/bensharkey).

## Author Contributions

BNLS conducted planning of infrared observations, data processing and analysis, and drafting the manuscript. VR initiated characterization campaign in 2016 and led the planning and coordinating of the overall observing campaign, as well as assisting in the compositional analysis and manuscript drafting. He is also the PI of the NASA grant that funded BNLS. RM conducted dynamics calculations, including figure preparation, and drafting of text. AT conducted observations and processing of LDT data and performed lightcurve analysis. OK conducted spectroscopic observations at LBT and assisted in planning and preparation of LBT programs, and processing of the visible spectrum. AC organized observing campaigns at LBT, including drafting observing proposals and assisting in data collection. BR assisted in collection and processing of LBT infrared data. JAS provided analysis of NEA comparisons. DT assisted in LBT data collection and planning. CV conducted observing planning and coordination of programs at LBT. All authors assisted in revising and editing the manuscript throughout the review process.

## Competing Interests

The authors declare no competing interests.

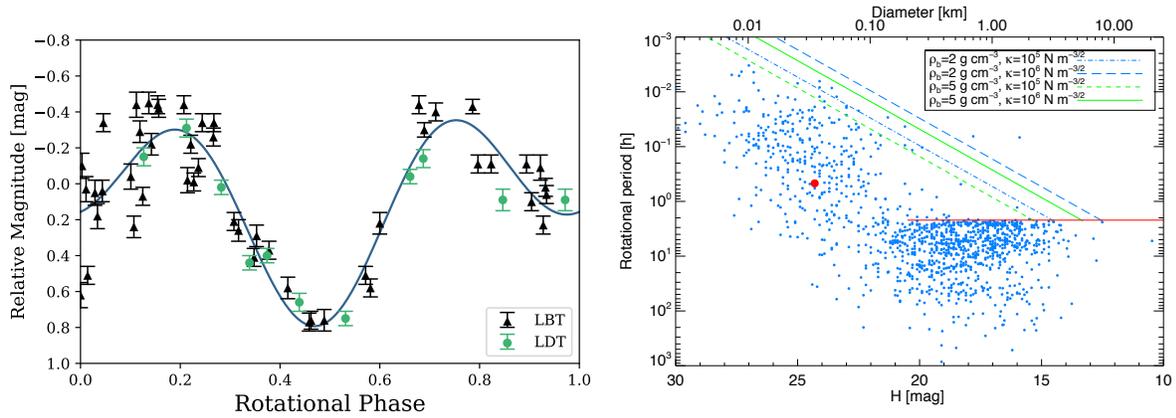

Figure 1. Kamo`oalewa lightcurve and Rotational Properties. Combined lightcurve of Kamo`oalewa from LBT and LDT, phased to a rotational period of 28.3 minutes (a), with comparison to the known sample of small near-Earth objects (b, Kamo`oalewa indicated as the large red point). We find that the observations are well described by a double-peaked lightcurve with asymmetric minima (blue curve, left panel) with an amplitude of 1.07±0.05 mag. Though it rotates faster than the 2.2 hour strengthless spin barrier seen with larger asteroids, when compared to similarly-sized NEOs, Kama`oalewa's spin period is typical. The diagonal curves in the right panel indicate limits of rotational breakup for asteroids given several assumed bulk densities $(\rho_b)$ and tensile strength coefficients $(\kappa)$. Translation from absolute magnitudes (H) to diameter [58] assumes a geometric albedo of 0.20. The lightcurve is shown with photometric uncertainties (random error) as the error bars.

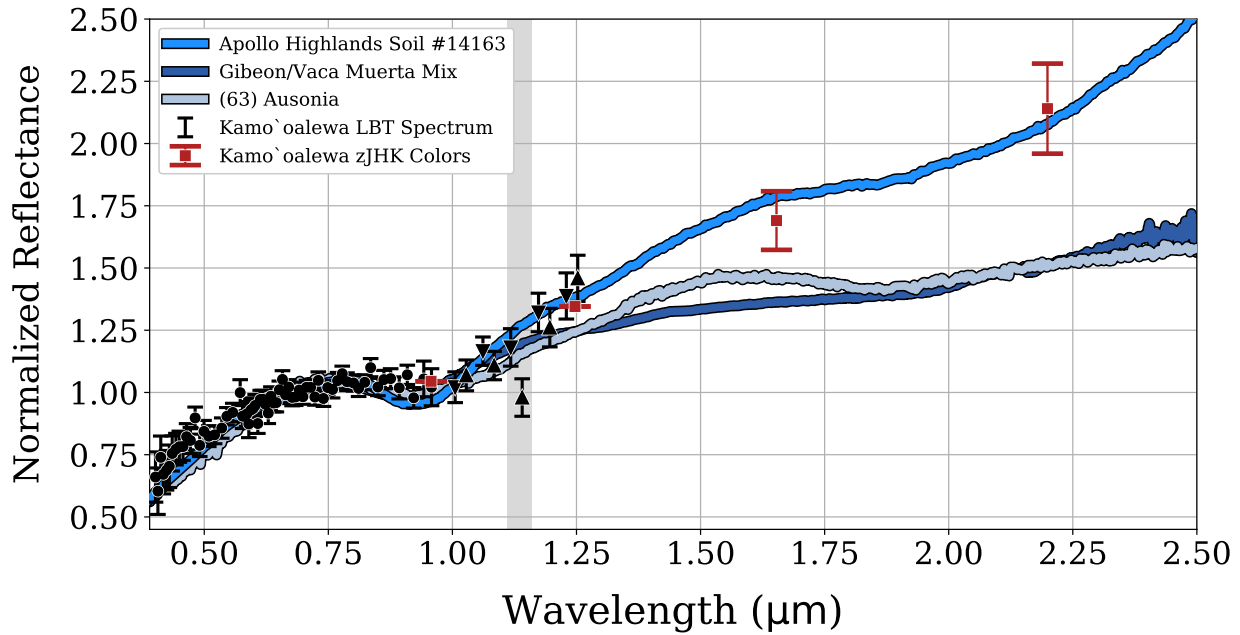

Figure 2. Kamo`oalewa's Reflectance Spectrum. Comparison of VNIR reflectance spectrum of Kamo`oalewa with that of a virtual mixture of the meteorites Gibeon and Vaca Muerta and the reflectance spectrum of lunar highlands sample #14163 (grain size 20-45 µm) returned from the Apollo 14 mission [59]. Black circles indicate the spectrum collected using the MODS instrument in 2017 (0.4-0.95µm), and black triangles indicate the infrared spectrum collected using the LUCI instrument in 2019 (0.95-1.25 µm). Our data shows a spectrum differing strongly from reddened silicate-rich asteroid spectra, exemplified by the Sw-type (63) Ausonia [21-22]. The steeply red-sloped spectrum we observe is consistent with a highly space-weathered silicate surface, similar to that of lunar samples. The gray shaded region indicates wavelengths with time-variable telluric features that can introduce artifacts in the data. Spectra were normalized to unity at 0.7 µm. The error bars shown for the spectrum are the photometric uncertainties at the spectral resolution as shown. The error bars for the zJHK colors represent the errors in the color ratio measurements (independent measurements were made of z/H, J/H, J/K).

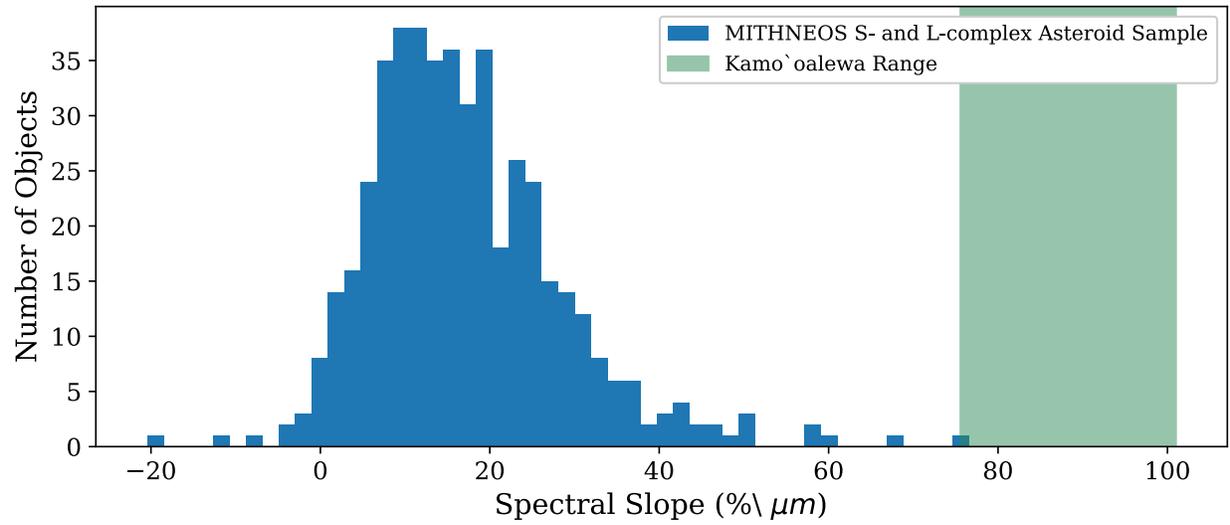

Figure 3. Comparison of Kamo`oalewa's Spectral Slope with Typical NEAs. Histogram (50 equal-width bins) of S- and L-type asteroid spectral slopes from the MITHNEOS spectral survey of NEAs [26]. Our measured spectral slope for Kamo`oalewa (green shaded region lies outside the typical range of 99.8% of the 470 S- and L-complex asteroids from the distribution shown in blue).

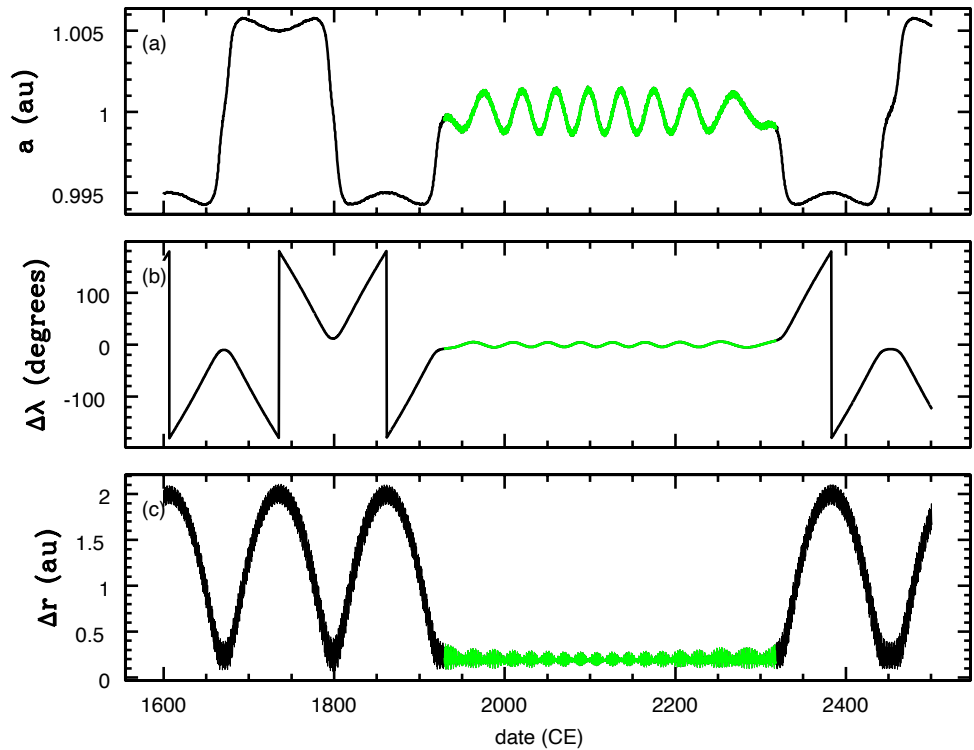

Figure 4. Kamo`oalewa's Current Orbital State.
(a) Orbital variation of the semi-major axis of Kamo`oalewa over ± 500 years; (b) time variation of the mean longitude relative to Earth's; (c) time variation of Kamo`oalewa's geocentric distance. The green track indicates the quasi-satellite state when the asteroid's semi-major axis as well as mean longitude remains very close to Earth's; the black track indicates the horseshoe state when the asteroid's semi-major axis alternates between superior and inferior to Earth's as its mean longitude approaches and recedes from Earth's. (Based on the DE430/431 planetary ephemeris [60], data provided by the Jet Propulsion Laboratory's HORIZONS on-line solar system data and ephemeris computation service; retrieved July 3, 2021.)

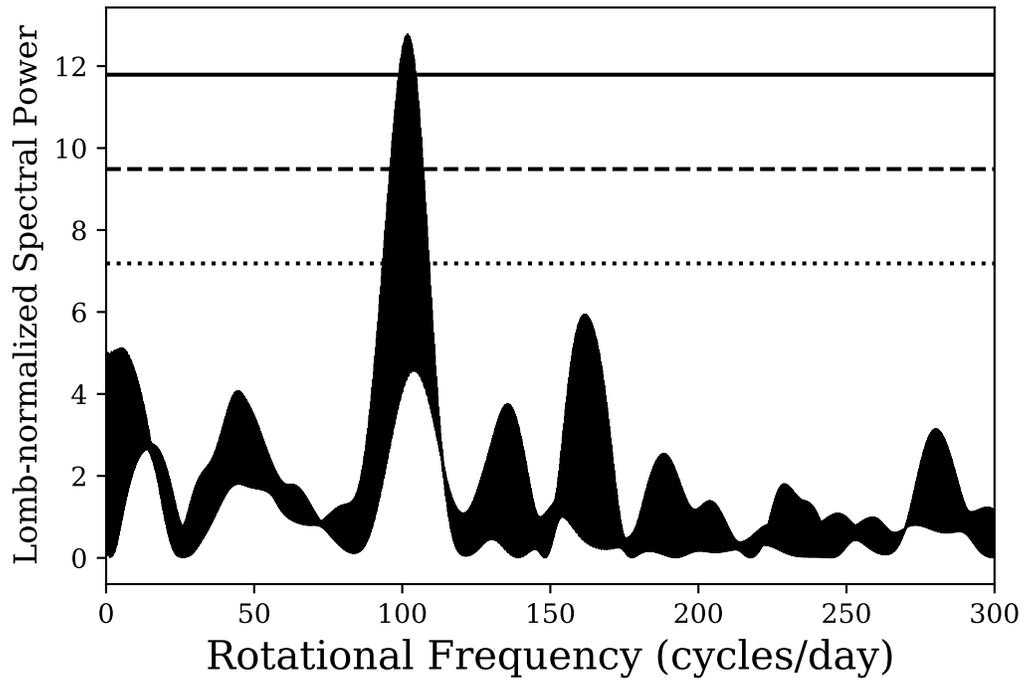

Figure 5. Lomb-Scargle periodogram from combined LBT and LDT photometry. We find that the observations are well described by a double-peaked lightcurve with asymmetric minima, corresponding to a rotational period twice that of the single-peaked solution shown in the periodogram. Dotted, dashed, and solid lines plotted on the periodogram indicate confidence intervals of 99.9%, 99.0%, and 90.0%, respectively.

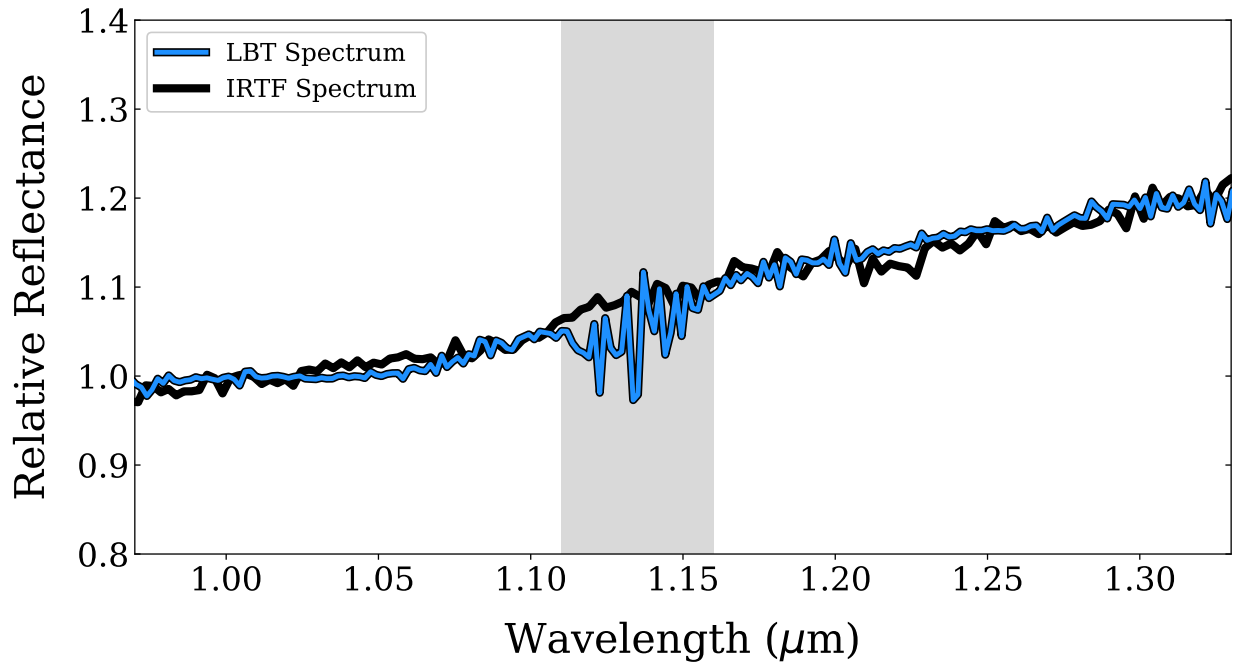

Figure 6. Data Processing Validation with NASA Infrared Telescope Facility. Observations of bright main-belt asteroid (26) Proserpina used to validate our processing methods. The asteroid was first observed using the SpeX instrument at the NASA Infrared Telescope Facility (IRTF) [61] and processed with SpexTool [62]. The object was then observed at LBT with the same instrument configuration as used to observe Kamo`oalewa and is shown binned to the resolution of the SpeX data. The gray shaded region indicates the same wavelength region as indicated in Figure 2, illustrating the effects of noise from telluric absorption features in our observing mode. Photometric errors are plotted but are not visible at this scale for this bright calibration target.